  \documentstyle[12pt]{article}
  
\catcode`@=11
\def\secteqno{\@addtoreset{equation}{section}%
\def\theequation{\thesection.\arabic{equation}}}
\catcode`@=12
\secteqno
\newcommand{\be}{\begin{equation}}
\newcommand{\ee}{\end{equation}}
\newcommand{\bea}{\begin{eqnarray}}
\newcommand{\eea}{\end{eqnarray}}
\newcommand{\bref}[1]{(\ref{#1})}
\newcommand{\ep}{\epsilon}

\newcommand{\A}{\alpha} \newcommand{\B}{\beta}
 \newcommand{\D}{\delta}
         \newcommand{\lam}{\lambda}
\newcommand{\r}{\rho}           \newcommand{\s}{\sigma}

\newcommand{\h}{\eta}

\def\pa{\partial}

\def\CL{{\cal L}}

\def\nb{\nabla}
\def\srg{\sqrt{\gamma}}

\newcommand{\nn}{\nonumber}

\def\t{\tilde}

\def\l{{\ell}}
\begin{document}
          \hfill TOHO-FP-9861

	  \hfill YITP-98-33

\vskip 20mm 
\begin{center} 
{\bf \Large Electric-Magnetic Duality Rotations and Invariance of Actions} 

\vskip 10mm
{\large Yuji\ Igarashi, Katsumi\ Itoh$^a$\footnote{Permanent address: Faculty of Education, Niigata University, Niigata 950-21,
Japan} and Kiyoshi\ Kamimura$^b$}\par 

\medskip
{\it 
Faculty of Education, Niigata University, Niigata 950-21, Japan\\
$^a$ Yukawa Institute for Theoretical Physics, Kyoto University\\Kyoto
606-01, Japan\\
$^b$ Department of Physics, Toho University, \ Funabashi\ 274, Japan\\
}

\medskip
\date{\today}
\end{center}
\vskip 10mm
\begin{abstract}

For D=4 theories of a single U(1) gauge field strength coupled to
gravity and matters, we show that the electric-magnetic duality can be
formulated as an invariance of the actions.  The symmetry is associated
with duality rotation acting directly on the gauge field. The rotation
is constructed in flat space, and an extension to curved spaces is also
given. It is non-local and non-covariant, yet generates off-shell
extended transformation of the field strength.  The algebraic condition
of Gaillard and Zumino turns out to be a necessary and sufficient
condition for the invariance of actions. It may be used as a guiding
principle in constructing self-dual actions in string and field
theories.

\end{abstract}

\noindent
{\it PACS:} 11.15.-q; 11.10.Ef\par\noindent
{\it Keywords:} Duality in Gauge Field Theories; D-branes; String Duality

\newpage
\setcounter{page}{1}
\setcounter{footnote}{0}
\parskip=7pt
\section{Introduction} 

Duality symmetries which relate strong couplings to weak couplings have
been considered as a key concept in understanding non-perturbative
aspects of string and field theories.  The D=4 electric-magnetic duality
is the prototype of the symmetries. It is associated with a rotation
between electric and magnetic fields, described by an infinitesimal
transformation of field strength to be called as the {\it
F-transformation} in this paper.  In the presence of matter fields, the
rotation should be accompanied, in general, by their appropriate
transformations.  The duality is not known to be manifest in the action
but only as a symmetry of equations of motion (EOM). In view of
increasing recognition of its importance, it is certainly desirable to
incorporate this symmetry directly into the action.

The purpose of this paper is to show that the duality can be formulated
as a pseudo-invariance\footnote{When an action is not exactly invariant
but transforms by a surface term, it is said to be pseudo-invariant.} of
a generic action for interacting U(1) gauge field strength with gravity
and matters.  There are two crucial issues in this formulation.  The
first is to realize the importance of the Gaillard-Zumino (GZ)
condition\cite{GaillardZumino}\cite{GaillardZumino2} on the
action.\footnote{This condition was discussed also by Gibbons and
Rasheed\cite{GibbonsRasheed} in the absence of matter fields.}  It was
originally found as a condition for the invariance of the stationary
surface.  Obviously it is the necessary condition for the
pseudo-invariance of the action. We show that the GZ condition is at the
same time a sufficient condition: if the action satisfies the GZ
condition, it becomes pseudo-invariant.  One needs to specify the
transformations of matter fields so as to be consistent with the
condition. Therefore, this algebraic relation provides us with a
criterion in construction of self-dual actions in string and field
theories. The second is to consider a duality transformation of the
fundamental dynamical variable, the gauge field, rather than that of the
field strength.  It is a natural extension of the one given by Deser and
Teitelboim\cite{DeserTeitelboim}, referred to as {\it A-transformation}
in this paper, being non-local and non-covariant, yet allows us to
formulate the duality as an invariance of the action. The
A-transformation is first constructed in flat space, and extended to
generic curved spaces under some assumptions on the inverse Laplacian
operator for D=3 spatial vectors.  Our formulation given here should be
compared with models with dual gauge fields such as the Schwarz-Sen
model\cite{SchwarzSen}, and its covariant version of PST\cite{Tonin}.
Such approaches of doubling gauge fields may be used to avoid above
mentioned non-locality and the sacrifice of D=4 manifest covariance. It
seems however that general criterion for constructing self-dual actions
with matter fields such as the GZ condition is not known in the
approaches with dual gauge fields.

We stress that since the F-transformation mixes the Bianchi identities
with the EOM, it is only sensible for the on-shell gauge field
configurations. Unlike this, the A-transformation is constructed so as to
admit off-shell field configurations, and to induce the
F-transformation on the mass shell.  The action becomes
pseudo-invariant under the A-transformation, if it obeys the GZ
condition.  As a non-trivial application of the formulation given here,
we show in a subsequent publication\cite{IIK} that the super D3-brane
action satisfies the GZ condition, and therefore is exactly self-dual.
A new feature found in that work is that fermionic co-ordinate of the
brane should transform while their bosonic super partner are left
invariant under the duality transformation.  This proof of self-duality
can be done without resort to any semi-classical approximations in
contrast to earlier works.

In our formulation based on the A-transformation, pseudo-invariance of
the action in the Lagrangian formalism implies an invariance of the
Hamiltonian in the canonical formalism. It has been discussed that the
Maxwell action is not pseudo-invariant under a finite duality
transformation, while the Hamiltonian is invariant. The source of this
conundrum is the use of a finite F-transformation. In the
A-transformation, there is no such discrepancy between the Lagrangian
and Hamiltonian approaches.  In the latter, the A-transformation is
shown to be described as a canonical transformation. Therefore, the
criterion for self-duality in the canonical formalism is commutability
of the Hamiltonian with the generator of the canonical transformation.

This paper is organized as follows. The next section describes the
F-transformation and the GZ condition. In section 3, A-transformation in
the flat space-time is given. We discuss a curved space-time extension
of the A-transformation in section 4. We also show that the
A-transformation is described as a canonical transformation. Summary and
discussion is given in section 5.  In Appendix A, we show that the GZ
condition is an algebraic relation which is not restricted to the
on-shell fields. Some formulae used in the D=3 covariant calculations
are summarized in Appendices B and C.


\section{F-transformation and the Gaillard-Zumino Condition} 

We begin with a brief review of duality transformation on field strength
and the GZ condition.  It was
shown\cite{GaillardZumino}\cite{GaillardZumino2} that the maximal
duality group associated with infinitesimal transformations of $n$ U(1)
field strengths is Sp(2n,R). This non-compact group is realized only when
there are appropriate scalar fields in the theory.  In the absence of
them, the relevant group is its maximal compact subgroup,
U(n). Conversely, the U(n) duality can be lifted to the non-compact
Sp(2n,R) duality by including some scalar fields, as discussed by
Gibbons and Rasheed\cite{GibbonsRasheed2} and Gaillard and
Zumino\cite{GaillardZumino2}.  We consider here for simplicity an SO(2)
duality for a single gauge field strength $F_{\mu\nu}$ coupled with D=4
metric $g_{\mu\nu}$ and matter fields $\Phi^A$.  Given a generic
Lagrangian density $\CL(F_{\mu\nu}, g_{\mu\nu}, \Phi^A)=\sqrt{-g}
L(F_{\mu\nu}, g_{\mu\nu}, \Phi^A)$, one may define the constitutive
relation by
\bea
\t K^{\mu\nu}~=~\frac{\pa L}{\pa F_{\mu\nu}},~~~~~~~~~~
\frac{\pa F_{\A\B}}{\pa F_{\mu\nu}}~=(\D^\mu_\A~\D^\nu_\B~-~
\D^\mu_\B~\D^\nu_\A),
\label{defKt}
\eea
where the Hodge dual components\footnote{Here we use the following
convention: $\eta^{\mu\nu\r\s}$ denotes the covariantly constant
anti-symmetric tensor with indices raised and lowered using the metric
$g_{\mu\nu}$ whose signature is $(-+++)$. We also use the tensor
densities $\ep^{\mu\nu\r\s}$ and $\ep_{\mu\nu\r\s}$ with weight $-1$ and
$1$.  In terms of $g=\det g_{\mu\nu}$, they are defined by $\ep^{\mu\nu\r\s}
=\sqrt{-g}\eta^{\mu\nu\r\s}$ and $\eta_{\mu\nu\r\s} =\sqrt{-g}
\ep_{\mu\nu\r\s}$, normalized as $\ep^{0123}~=~-\ep_{0123}~=~1$.} for the
anti-symmetric tensor $K_{\mu\nu}$ are given by
\bea 
\t K_{\mu\nu}~=~\frac{1}{2} \eta_{\mu\nu}~^{\r\s}~K_{\r\s},~~~~~~
K_{\mu\nu}~=~-~\frac{1}{2} \eta_{\mu\nu}~^{\r\s}~\t K_{\r\s},~~~~~~ 
{\tilde {\t K}}_{\mu\nu}~ =~-~K_{\mu\nu}.  
\eea
In terms of $F_{\mu\nu}$ and $K_{\mu\nu}$, the EOM for gauge field and the
Bianchi identity read
\bea
{\pa}_{\mu}(\sqrt{-g} \t K^{\mu\nu} )=\frac12\ep^{\mu\nu\r\s}{\pa}_{\mu}
  K_{\r\s}  =0, \nonumber\\
{\pa}_{\mu}(\sqrt{-g} \t F^{\mu\nu} )=\frac12\ep^{\mu\nu\r\s}{\pa}_{\mu}
  F_{\r\s}  =0~.
\label{eombi}
\eea
The SO(2) duality has been postulated as a symmetry of the EOM: its
infinitesimal form is
\bea
\D F~=~\lam~K,~~~~~\D K~=~-~\lam~F,
\label{DFK}
\eea
associated with 
\bea
\D\Phi^A~=~\xi^A(\Phi),~~~~~~\D g_{\mu\nu}=0,
\label{DT}
\eea
where $\xi^A (\Phi)$ should depend neither derivatives of the fields,
$\pa_{\mu} \Phi^A =\Phi_{\mu}^{A}$, nor other fields, $g_{\mu\nu}$ and
$F_{\mu\nu}$.  Note that $F_{\mu\nu}$ and $K_{\mu\nu}$ are not
independent one another but non-linearly related in general.  Therefore,
under \bref{DFK}, only a restricted class of actions will make the
relation \bref{defKt} invariant. This constraint on actions is the GZ
condition, which may be obtained out of some equations given in
refs.\cite{GaillardZumino}\cite{GaillardZumino2}.  For SO(2) duality, it
is given by
\bea
\frac{\lam}{4}(~F~\t F~+~K~\t K)~+~\D_\Phi L~=~0,
\label{u1duality}
\eea
where  $F~\t F= F_{\mu\nu}{\t F}^{\mu\nu}$.

The GZ condition can be derived by requiring: \\
(1) invariance of the constitutive relation \bref{defKt}; \\
(2) covariance for the EOM of matter
fields\cite{GaillardZumino}\cite{GaillardZumino2},
\bea
\D~[{\cal L}]_{A} = - \frac{\pa \xi^B}{\pa \Phi^A}~[{\cal L}]_{B},
~~~~~~~~[{\cal L}]_{A}=\frac{\pa {\cal L}}{\pa
\Phi^A}-\pa_{\mu}\frac{\pa {\cal L}}
{\pa \Phi^{A}_{\mu}};
\label{meom}
\eea  
(3) invariance of the EOM for the metric $g_{\mu\nu}$, which implies
invariance of the energy-momentum tensor
\bea \D~T^{\mu\nu}~=~0,~~~~~~~~~
\frac{\D S}{\D g_{\mu\nu}}~=~-\frac{\sqrt{-g}}{2}T^{\mu\nu}~
\eea
under the duality transformation \bref{DFK} and
\bref{DT}. 

First, one finds the constitutive relation \bref{defKt} to
transform as
\bea
\D {\t K}= \frac{1}{2}\frac{\pa \t K}{\pa F}~\lam K  ~+~
\frac{\pa}{\pa F} \D_{\Phi} L 
= -\lam {\t F} ~+~\frac{\pa}{\pa F}~\beta,
\label{DtK1}
\eea
where
\bea
\beta~\equiv~ \frac{\lam}{4}\left({\t
K}K + {\t F}F\right) + \D_{\Phi} L.  
\label{beta}
\eea
As shown in Appendix A, the EOMs for the matter fields change as
\bea
\D ~[{\cal L}]_{A} =  - \frac{\pa \xi^B}{\pa \Phi^A}~[{\cal L}]_{A}~+~
[\sqrt{-g}~\beta]_{A}.
\label{GZder4}
\eea
We also find 
\bea
\D ~\left(\frac{\D S}{\D g_{\mu\nu}(x)}\right) &=&  
\frac{1}{2} \lambda \int d^4 y \left(K_{\rho\sigma}(y)\frac{\D}{\D g_{\mu\nu}(x)}~({\sqrt
-g}{\tilde K^{\rho\sigma}})(y)
+\frac{\D}{\D g_{\mu\nu}(x)}~\delta_{\Phi}{\cal L}(y)\right)
\nn\\
&=&  \frac{\D}{\D g_{\mu\nu}(x)}~\int d^4 y (\sqrt{-g}~\beta)(y).
\label{GZder44}
\eea
The above three conditions 1) $\sim$ 3) require that $\sqrt{-g}\beta$
should be a constant.  Moreover, in order for the relation \bref{beta}
to have a consistent power series expansion in terms of
$(F,~\Phi^A,~\Phi^{A}_{\mu})$, the constant should be zero, so that
$\beta=0$.  This leads to the GZ condition. 

As mentioned in the previous section, the F-transformation is only
sensible for on-shell gauge field configurations.  So one might think,
from the above derivation, that the GZ condition holds only for on-shell
fields.  However, it is not so: the GZ condition is an algebraic
relation that is free from on-shell conditions in spite of its
derivation. In Appendix A, we show it explicitly in connection with
A-transformation which generates off-shell extended transformations of
the field strength.

Consider the on-shell gauge field configurations appeared in the
F-transformation, and suppose that the matter fields are absent.  Then
the Lagrangian density transforms by a term proportional to
$~\sqrt{-g}K~\t K$.  In the absence of matter fields, the change of the
Lagrangian density under the F-transformation is $~\sqrt{-g}K~\t K$ or
$~K_{(2)}^2~$ in terms of the differential two form $K_{(2)}$.  Since
the transformation is sensible only for the on-shell gauge
configuration, i.e., $d~K_{(2)}=0$, $K_{(2)}^2$ becomes exact if the
cohomology of $K_{(2)}^2$ is trivial.  In this sence, the
pseudo-invariance of the action under the F-transformation trivially
follows. What is non-trivial is that even in the presence of the
(off-shell) matter fields, the action still remains pseudo-invariant, if
it obeys the GZ condition.  This is why Gaillard and
Zumino\cite{GaillardZumino}\cite{GaillardZumino2} were able to construct
a conserved charge associated with the F-transformation.  In any case,
however, the restriction on the on-shell variation for the gauge field
in the F-transformation would be undesirable.  This tempts us into
considering the A-transformation.  We consider it below first in flat
space and then in curved spaces.


\section{A-transformation in flat space}

In order to see the essential issue of duality rotation on the gauge
field as simple as possible, and to avoid technical complications
appeared in curved spaces, we first consider the A-transformation in
flat space\cite{DeserTeitelboim}.  We construct it in such a way that:

1) it allows to include off-shell field configurations and produces the
F-transformation on the mass-shell;

2) a generic action becomes pseudo-invariant if the GZ condition is
satisfied.

For SO(2) duality, the A-transformation which satisfies the above
requirements is given in a non-covariant (3+1) decomposition by
\bea
\D A_i&=&-~\lam~\ep_{ijk}~{\pa^{-2}}\pa^j~{\t K^{0k}},~~~~(i,j,k=1,2,3)
\nn\\
\D A_0&=&-\frac{\lam}{2} ~{\pa^{-2}}\pa^k
\left(\ep_{ijk}~\t K^{ij}\right),
\label{transA}
\eea 
where $\t K^{\mu\nu}$ is defined by \bref{defKt} in flat space and is in
general a non-linear function of $(F,~\Phi^A,~\pa\Phi^A)$. 
In relation to the inverse Laplacian operator ${\pa^{-2}}$ we assume
that: a) non-trivial kernel does not exist; b) it commutes with
derivatives, $\pa_\mu{\pa^{-2}}={\pa^{-2}}\pa_\mu $, and allows the
partial integration, $f{\pa^{-2}}g = ({\pa^{-2}}f)~g+({\rm
surface~term})$.

The A-transformation \bref{transA} generates changes in the field strength
\bea
\D F_{ij} &=& \lam~\ep_{ijk}~\t K^{0k}_\perp
\nn\\
&=& \lam~K_{ij} ~+~\lam~\ep_{ijk}~{\pa^{-2}}{\pa^k}~[~
\pa_\mu \t K^{\mu 0}~],
\nn\\
\D F_{0i} &=& \pa_0\Bigl(-\lam~\ep_{ijk}~{\pa^{-2}}{\pa^j}\t K^{0k}\Bigr)~-~
\pa_i\Bigl(-~\frac{\lam}{2}~{\pa^{-2}}{\pa^j}(\ep_{jk\l}~\t K^{k\l})\Bigr)
\nn\\
&=& \lam~K_{0i} ~-~\lam~\ep_{ijk}~{\pa^{-2}}{\pa^j}~[~\pa_\mu \t K^{\mu k}~],
\label{transF}
\eea
where $[~~]$ indicates contributions which vanish on the mass-shell. The
use of EOM reduces \bref{transF} to $\D F_{\mu\nu}~=~\lam~K_{\mu\nu}$. As shown
in Appendix A, the GZ condition \bref{u1duality} and \bref{transF} lead
to
\bea
\D \t K^{\mu\nu} = -{\lam}\t F^{\mu\nu}~+~
\frac12\frac{\pa \t K^{\r\s}}{\pa F_{\mu\nu}}~[EOM]_{\r\s},
\label{deltK}
\eea
where $[EOM]_{\r\s}$ are the components of EOM in \bref{transF}. This
gives rise to $\D \t K^{\mu\nu}=-~\lam~\t F^{\mu\nu}$ on the mass shell
again. One finds that the A-transformation \bref{transA} satisfies the
requirement 1).

We next consider the change of the Lagrangian under \bref{transA}:
\bea
\D L&=&\frac12~\frac{\pa L}{\pa F_{\mu\nu}}\D F_{\mu\nu}~+~\D_\Phi L~
\nn\\
&=&\t K^{0i}\left(\lam~K_{0i} -~\lam~\ep_{ijk}~{\pa^{-2}}{\pa^j}
[\pa_\mu \t K^{\mu k}]\right)
\nn\\
&~&+~\frac12~\t K^{ij}
\left( \lam~K_{ij} ~+~\lam~\ep_{ijk}~{\pa^{-2}}{\pa^k}[\pa_\mu \t K^{\mu
0}]\right)~+~\D_\Phi L
\nn\\
&=& \frac{\lam}{2}~\t K K~+~\D_\Phi L~-~
\lam~\t K^{0i}~\ep_{ijk}~{\pa^{-2}}{\pa^j}
[~\pa_\mu \t K^{\mu k}~]
\nn\\
&~&+~
\frac12~\lam~\t K^{ij}\ep_{ijk}~{\pa^{-2}}{\pa^k}[~\pa_\mu \t K^{\mu 0}~].
\label{DL12}
\eea
The last two terms proportional to the EOM become
\bea
-~\frac{\lam}{4}~\t K K~+~\pa_0\left(-\frac{\lam}{2}~\t K^{0i}
\ep_{ijk}~{\pa^j}{\pa^{-2}}~\t K^{0k}\right)
\eea
up to a total spatial derivative.  Imposing the GZ condition
\bref{u1duality}, the Lagrangian is shown to transform by a total D=4
divergence:
\bea
\D L&=&\pa_0\left(-\frac{\lam}{2}~\t K^{0i}\ep_{ijk}~{\pa^{-2}}{\pa^j}~\t K^{0k}~-~
\frac{\lam}{2}~A_{i}\ep^{ijk}{\pa_j}~A_{k}~\right)~+~\pa_i U^i
\nn\\
&\equiv&~\pa_\mu~ U^\mu,
\label{DL14}
\eea
for some $U^i$.  One confirms that the action becomes pseudo-invariant
for off-shell gauge field under the A-transformation, if it obeys the GZ 
condition.

The pseudo-invariance of the action leads to the conserved current,
$\pa_{\mu} j^\mu =0$.  
In particular, the charge
density is given by
\bea
j^0&=&\t K^{0i}\D A_i~+~
\frac{\pa L}{\pa \dot\Phi^A}\D \Phi^A~-~U^0
\nn\\
&=&-~\frac{\lam}{2}~\t K^{0i}~\ep_{ijk}{\pa^{-2}}{\pa^j}~{\t K^{0k}}~+~
\frac{\lam}{2}~A_{i}\ep^{ijk}{\pa_j}~A_{k}~+~
\frac{\pa L}{\pa \dot\Phi^A}\D \Phi^A.
\label{consc}
\eea

The discussion given here in the flat space-time shows that the duality
symmetry can be formulated as a pseudo-invariance of the action if the
GZ condition is fulfilled.  Therefore the GZ condition, though
originally derived as a necessary condition, is a sufficient condition
for the pseudo-invariance for the action as well.  It is recognized to
be a criterion for the pseudo-invariance of the action.  We now consider
an extension of the formulation in curved spaces.

\section{Extension to curved space}

In this section, we construct a curved-space version of the
A-transformation, where covariance with respect to D=3 subspace is
maintained. Basic notations and formulae needed to perform D=3
covariant computations are summarized in Appendix B.

One of the basic operators in our formalism is a curved space extension
of the Laplacian operator, $(\t\Delta)_{j}^{~i}$, which maps a vector
$T_i$ into a vector $(\t\Delta)_{j}^{~i}T_{i}$ . It is given by
\bea
(\t\Delta)_{j}^{~i}~=~\Delta \delta_{j}^{~i}-R_{j}^{~i},~~~~~~~~~~
\Delta~=~\nb^j \nb_j. 
\label{MLap}
\eea
where $R_{j}^{~i}$ is the Ricci tensor. We assume that boundary
conditions can be arranged so that the Laplacian operator has no
non-trivial kernel, and its inverse, $(\t\Delta^{-1})_i^{~j}$, is
well-defined.

In terms of this inverse operator, the transformation for spatial
components of the gauge field is given by
\bea
\D A_\l~=~
\lam~(\t\Delta^{-1})_\l^{~k}\nb^j\ep_{jkm}~{\t {\cal K}}^{0m}
~=~D^{-1}_{\l m}~\left(\lam~\frac{{\t {\cal K}}^{0m}}{\srg}\right)\equiv
 \lam Z_l,
\label{transAL}
\eea
where ${\t{\cal K}}^{0i}$ is a D=3 vector
density defined via D=4 tensor density
\bea
\t {\cal K}^{\mu\nu}~=\sqrt{-g} {\t K}^{\mu\nu}.
\label{defcKt}
\eea
Note that $\t{\cal K}^{0i}$ divided by $\srg~=\sqrt{\det g_{ij}}$
becomes a D=3 vector. In \bref{transAL}, $D^{-1}_{i\l}$ is a tensor
operator acting on a vector,
\bea
D^{-1}_{i\l}\equiv(\t\Delta^{-1})_i^{~k}\nb^j\h_{jk\l}
~=~\h_{ijk}\nb^k(\t\Delta^{-1})^j_{~\l},
\label{defDI}
\eea
where $\h_{jk\l}=\ep_{jk\l}~{\srg~}$ is the covariantly constant
anti-symmetric tensor. $D^{-1}_{i\l}$ is the inverse of 
\bea
D^{jk}~=~\h^{j\l k}~\nb_\l~=~\nb_\l~\h^{j\l k},
\label{defD}
\eea
in a projected space,
\bea
D^{-1}_{im}~D^{mk}&=&O_i^{~k}(\nb),~~~~~~~~
D^{im}~D^{-1}_{mk}~=~O^i_{~k}(\nb),
\nn\\
O_i^{~k}(\nb)&=&\D_i^{~k}~-~\nb_i(\Delta^{-1})\nb^k.
\label{FM1}
\eea
Note that the operator $O_i^{~k}(\nb)$ projects out any longitudinal
component defined with the covariant derivative $\nb_i$.  These tensor
operators satisfy the transverse condition
\bea
\nb^k D^{-1}_{k\l}T^\l~=~\nb_k D^{k\l}T_\l~=~0.
\label{FM2}
\eea
Computing $\nb^i \D A_i$ and $D^{ij} \D A_j$, one obtains the
transverse condition
\bea
\nb^i \D A_i~=~0,
\label{tvc}
\eea
and the transformation of the magnetic field
\bea
\D F_{ij}~=~\lam~\ep_{ijk}\t {\cal K}^{0k}_\perp,
\label{traFjk}
\eea
where
\bea
\t {\cal K}^{0k}_\perp&=&\t {\cal K}^{0k}~-~
{\srg~}\nb^k(\Delta^{-1})~\left(\nb_m~\frac{{\cal K}^{0m}}{\srg~}\right)
\nn\\
&=&\ep^{k\l m}\pa_\l Z_m.
\label{defZL}
\eea
The transformation of the time component is given by 
\bea
\D A_0~=~-~\frac{\lam}{2}~\left[
\nb^\l(\t\Delta^{-1})_{\l}^{~k}(\ep_{ijk}~\t {\cal K}^{ij})\right]~+~
\lam [(\Delta^{-1})~\nb^k ~(\pa_0 Z_k)].
\label{transA0}
\eea
It leads to 
\bea
\D F_{0i}&=& \pa_0 \D A_{i}-\pa_i \D A_0 
\nn\\
&=&{\lam}~K_{0i}~+~{\lam} D^{-1}_{ij} 
\left(\frac{\pa_{\mu}{\t {\cal K}}^{\mu j}}{\srg}\right) + \lam 
 O_i^{~k}
\pa_0\left( \frac{D^{-1}_{kj}}{\srg}\right)~\t {\cal K}^{0j},
\label{delf0i}
\eea
where $O^i_{~k}$ is the transverse projection in \bref{FM1}. As shown
in Appendix C, the last two terms vanish on the mass shell, leaving the desired
first term.

We now consider the variation of the Lagrangian density given by
\bea
\D {\cal L}&=&{\t {\cal K}}^{0i}\D F_{0i}~+~\frac12~{\t {\cal
K}}^{ij}\D F_{ij}~+~\D_\Phi {\cal L}
\nn\\
&=&{\t {\cal K}}^{0i}~\pa_0\D A_i~-~{\t {\cal K}}^{0i}~\pa_i \D A_0~+~
\frac12~{\t {\cal K}}^{ij}(~\lam~\ep_{ijk}{\t {\cal
K}}^{0k}_\perp)~+~\D_\Phi {\cal L}.
\label{DL1}
\eea
One introduces here an equality up to a spatial total derivative,
$\sim$: $A\sim B \Leftrightarrow A~=~B~+~\pa_{i}V^{i}$ for some $V^{i}$. 
Then, one rewrites the first two terms
\bea
{\t {\cal K}}^{0i}~\pa_t\D A_i~-~{\t {\cal K}}^{0i}~\pa_i \D A_0
&\sim& \pa_0 \left[\frac{\lam}{2} Z_i\ep^{ijk}\pa_j Z_k\right]
\nn\\
&~&+
~(\pa_i~{\t {\cal K}}^{0i})~\left[\D A_0 - (\Delta^{-1})~\nb^k ~(\pa_0\D
A_k)\right],
\label{ftt}
\eea
and the third term 
\bea
\frac12{\t {\cal K}}^{ij}~\D F_{ij}&=&
\frac{\lam}{2}{\t {\cal K}}^{ij}\ep_{ijk}{\t {\cal K}}^{0k}_\perp
\nn\\
&=&\frac{\lam}{2}{\t {\cal K}}^{ij}\ep_{ijk}{\t {\cal K}}^{0k}~-~
\frac{\lam}{2}{\t {\cal K}}^{ij}~\ep_{ijk}{\srg~}~(\t\Delta^{-1})^k_{~\l}~
\nb^\l\left(\nb_m~\frac{{\t {\cal K}}^{0m}}{\srg~}\right)
\nn\\
&\sim&+\frac{\lam}{4}{\t {\cal K}}^{\mu\nu}~ K_{\mu\nu}~+~
\frac{\lam}{2}~\left[
\nb^\l(\t\Delta^{-1})_{\l}^{~k}(\ep_{ijk}~{\t {\cal K}}^{ij})\right]~
(\pa_m~{{\t {\cal K}}^{0m}}).
\label{ttt}
\eea
Substituting $\D A_0$ given in \bref{transA0} and using the GZ condition, 
one finds
\bea
\D {\cal L}~&=&~
\pa_0~\left[\frac{\lam}{2} Z_i\ep^{ijk}\pa_j Z_k\right]~+~
\frac{\lam}{4}{\t {\cal K}}^{\mu\nu}~ K_{\mu\nu}~+~
\D_\Phi {\cal L} + \pa_i U^i
\nn\\
&=& 
\pa_0~\left[\frac{\lam}{2} Z_i\ep^{ijk}\pa_j Z_k~-~
\frac{\lam}{2}\left(~A_i~\ep^{ijk}~\pa_j A_k\right)\right] + \pa_i U^i
\nn\\
&=& ~\pa_{\mu}~U^{\mu},
\label{DL2}
\eea
for some $U^i$. This establishes that the generic action becomes
pseudo-invariant under the extended A-transformation in curved spaces.

We have considered above the duality symmetry in the Lagrangian
formalism. It is known that in the canonical formalism the symmetry can
be realized simply as an invariance of the Hamiltonian. We argue here
that the A-transformation can be described as a canonical
transformation.  Let $(A_{\mu},~\pi^{\mu})$ be canonically conjugate
variables, which are relevant to the duality transformation. From
\bref{DL2} the generator of the canonical transformation is given by
\bea {\cal W} =\frac{\lam}{2}~\int d^3 x ~{\srg}~\left[\frac{\pi^i}{\srg}~
{D^{-1}_{i\l}}~\frac{\pi^\l}{\srg}~+~ A_i~{D^{ij}} A_j~+~\pi^{0}~\D A_0\right],
\label{canogene} 
\eea 
with the identification, $\t{\cal K}^{0i} =\pi^i$. It generates the
transformations of the canonical variables 
\bea
\D A_i&=&~{\lam}~D^{-1}_{ij}~\frac{\pi^j}{\srg},
\nn\\
\D \pi^i&=&-{\lam}~\ep^{ijk}~\pa_j A_k,
\label{canotra}
\eea
which lead to the desired relations
\bea
\D B^i &=&\lam~\pi^i_\perp,
\nn\\
\D \pi^i_\perp &=&~-\lam~B^i,
\label{canotra2}
\eea
as well as the invariance of the Gauss law constraint, $\D
(\pa_i\pi^i)~=~0$.  Note that the last term proportional to $\D A_0$ in
\bref{canogene} contains ``velocity variables'' $\dot q$ in addition to
phase space variables $(q,~p)$. Such a term is interpreted as a
generalized canonical quantity considered in ref.\cite{Kamimura}. This
term may induce new contributions to the canonical transformation of
$\pi^{i}$ in \bref{canogene}. However since all of them are proportional
to the constraint $\pi^0 \approx 0$, they vanish on the constraint
surface, and therefore have not been included there.


\section{Summary and Discussion}

In this paper we have shown that electric-magnetic duality can be
formulated as a symmetry of actions.  It is based on an integrated
form of the duality transformation for the gauge field.  Our
reinterpretation of the GZ condition as the general criterion for the
invariance of the action reveals its importance as a guiding principle
in constructing self-dual actions for theories of interacting U(1) field
strength.  The question that one needs to address is that how 
the GZ condition restricts the form of the self-dual actions for given
matter field content. In the absence of matters, Gibbons and
Rasheed\cite{GibbonsRasheed}\cite{GibbonsRasheed2} showed that there are
as many self-dual actions as functions of a single variable. A related
question is how inclusion of supersymmetry eliminates such
ambiguity.

Although we have considered SO(2) duality, extension to U(n) duality for
$n$ interacting Maxwell fields will be straightforward. This duality
associated with the compact group may be lifted to the maximal
non-compact group Sp(2n,R), by introducing scalar fields which belong to
the coset space
Sp(2n,R)/U(n)\cite{GaillardZumino}\cite{GaillardZumino2}. They give a
non-linear realization of the Sp(2n,R).

It is natural to expect that if the action is pseudo-invariant under an
infinitesimal transformation for certain continuous symmetry, it is so
under a finite transformation for the symmetry, as long as the latter is
obtained by an integration of the former.  As for the duality symmetry,
this is of course the case in the A-transformation, but not in the
F-transformation: the pseudo-invariance under the F-transformation does
not lead to that under a finite F-transformation. Even for the
infinitesimal transformation, the action can be pseudo-invariant only
when it is expressed in terms of the gauge field as can be observed in
the following:
\bea \D S =
\lam G_{F}~S&=& 2~\lam \int d^4 x {\mbox{\boldmath $E$}}\cdot
{\mbox{\boldmath $B$}},~~~~~~ 
S=\int d^4 x \frac12\left(
{\mbox{\boldmath $E$}}^2 -{\mbox{\boldmath $B$}}^2\right),
\\
G_F&=&{\bf B}\frac{\D}{\D {\bf E}}~-~{\bf E}\frac{\D}{\D {\bf B}}.
\eea 
A finite transformation generates terms,
\bea 
e^{\lam G_{F}}~S~=S~+~ 2~\lam \int
d^4 x {\mbox{\boldmath $E$}}\cdot {\mbox{\boldmath $B$}}~+~
\lam^2~G_{F}~\int d^4 x {\mbox{\boldmath $E$}}\cdot {\mbox{\boldmath
$B$}} ~+~\cdots.
\eea 
The third term goes back to the Maxwell action if we perform $G_F$
operation, while we do not know how to do it once the integrant is expressed in
terms of the gauge field.  This observation would resolve a ``paradox''
that the Maxwell action is not pseudo-invariant under a finite duality
transformation, while it is so under an infinitesimal
transformation. The source of the ``paradox'' is the use of the
F-transformation. There is no such problem in the A-transformation.

An undesirable feature of our formulation is that locality as well as
manifest D=4 covariance are lost.  A way out of this drawback may be to
introduce a new gauge field whose curl gives the dual field strength $\t
K$. Actually, the Schwarz-Sen model\cite{SchwarzSen} and its covariant
version of PST\cite{Tonin} retain the locality and the covariance in the
duality rotation. In this approach, however, it is not clear if there
exists the general criterion for self-dual actions such as the GZ
condition.  To find it, if exists, is certainly important for
construction of manifestly self-dual actions in string and field
theories.  

\medskip\noindent
{\bf Acknowledgements}\par \medskip\par

K.I. would like to thank to the Yukawa Institute for its kind
hospitality extended to him.

\appendix
\section{ A derivation of the GZ condition as an algebraic relation}

In this appendix, we shall first derive \bref{GZder4} from the covariance of
the EOM for matter fields \bref{meom}, and then show that the GZ
condition is an algebraic relation sensible even off the mass shell. 

Under the duality transformation \bref{DFK} and \bref{DT}, the EOM
\bref{meom} transforms by
\bea
\D ~[{\cal L}]_{A} = \D\left(\frac{\pa {\cal L}}{\pa \Phi^A}\right)~-~
{\pa}_{\mu}~\D\left(\frac{\pa {\cal L}}{\pa \Phi^A_{\mu}}\right),
\label{a11}
\eea
where the first term on the r.h.s. reads
\bea
\D \left(\frac{\pa {\cal L}}{\pa \Phi^A}\right)= 
\D \Phi^{B}~\frac{{\pa}^2 {\cal
L}}{\pa \Phi^B \pa \Phi^A}~+\D \Phi^{B}_{\mu}~\frac{{\pa}^2 {\cal
L}}{\pa \Phi^B_{\mu} \pa \Phi^A}~+\frac12 \D F~\frac{{\pa}^2 {\cal
L}}{\pa F \pa \Phi^A}.
\label{a2}
\eea
Using \bref{defKt} and \bref{beta}, the last term in \bref{a2} is
expressed as
\bea
\frac12 \D F~\frac{{\pa}^2 {\cal L}}{\pa F
\pa \Phi^A}&=&\frac{\lam}{4}\sqrt{-g}~\frac{\pa}{\pa \Phi^A}(K \t K)
\nn\\
&=& \sqrt{-g}~\frac{\pa}{\pa \Phi^A}\left(\beta~-~\frac{\lam}{4}F {\t
F}~-~\D \Phi^B \frac{\pa L}{\pa \Phi^B}~-~\D \Phi^{B}_{\mu}\frac{\pa L}{\pa \Phi^B_{\mu}}\right).
\label{a3}
\eea
It follows from \bref{a2} and \bref{a3} that
\bea
\D \left(\frac{\pa {\cal L}}{\pa \Phi^A}\right)=\sqrt{-g}~\left(\frac{\pa
\beta}{\pa \Phi^A}~-~\frac{\pa \xi^B}{\pa \Phi^A}\frac{\pa L}{\pa
\Phi^B}~-~\Phi^C_{\mu}\frac{{\pa}^2 \xi^B}{\pa \Phi^C \pa
\Phi^A}\frac{\pa L}{\pa \Phi^B_{\mu}}\right).
\label{a4}
\eea
Likewise, one finds 
\bea
\D\left(\frac{\pa {\cal L}}{\pa \Phi^A_{\mu}}\right)=\sqrt{-g}~\left(
\frac{\pa\beta}{\pa \Phi^A_{\mu}}~-~\frac{\pa \xi^B}{\pa \Phi^A}\frac{\pa L}{\pa
\Phi^B_{\mu}}\right).
\label{a5}
\eea
Then, one obtains \bref{GZder4} from \bref{a11}, \bref{a4} and \bref{a5}.

Although the GZ condition has been obtained originally via the
 F-transformation, it is an algebraic relation that is free from the
on-shell conditions. To see this explicitly, we consider off-shell extended
transformations inferred by the A-transformation:
\bea
{\hat \D} F~&=&~\lam~K~+~[EOM] ,~~~~~{\hat \D} K~=~-~\lam~F~+~[EOM] 
\label{ofDFK}
\eea
where $[EOM]$ denotes terms proportional to the EOM of the gauge fields.
All such terms vanish on the mass shell, and we do not need their precise
form below. Covariance relation of the EOM for matters
may be replaced by
\bea
{\hat \D} ~[{\cal L}]_{A} = - \frac{\pa \xi^B}{\pa \Phi^A}~[{\cal L}]_{B}~+~[EOM].
\label{ofmeom}
\eea
One obtains then 
\bea
{\hat \D} ~{\t K}&=& -\lam ~{\t F} ~+~\frac{\pa}{\pa F} \left\{ \frac{\lam}{4}\left({\t
K}K + {\t F}F\right) + \D_{\Phi} L \right\}~+~\frac{1}{2}\frac{\pa \t
K}{\pa F}~ [EOM]
\nn\\
{\hat \D}~[{\cal L}]_{A}&=& - \frac{\pa \xi^B}{\pa \Phi^A}~[{\cal L}]_{B}~+\left[~\sqrt{-g} \left\{ \frac{\lam}{4}\left({\t
K}K + {\t F}F\right) + \D_{\Phi} L \right\}\right]_{A}
\nn\\
&~&+~\frac{1}{2}~[\sqrt{-g}{\t K}]_{A}~[EOM].
\label{ofcal}
\eea
Consistency of \bref{ofcal} with \bref{ofDFK} and \bref{ofmeom} leads to 
again the GZ condition:
\bea
\frac{\lam}{4}(~F~\t F~+~K~\t K)~+~\D_\Phi L~=~0.
\label{apu1duality}
\eea

\section{Basic formulae in D=3 covariant calculations}

We illustrate the Laplacian operator ${\t \Delta}^j_{~i}$ acting on a
vector by considering an integration of the curl 
\bea \pa_j A_k~-~\pa_k A_j = \nb_j A_k~-~\nb_k A_j~=~ \ep_{ijk} B^i,
\eea 
where $\nb_i$ is the covariant derivative with respect to D=3 spatial
metric $g_{ij}$ whose determinant is $\gamma$.  The anti-symmetric
tensor densities $\ep^{ijk}$ and $\ep_{ijk}$ are related with the D=4
tensor densities by $\ep^{0ijk}=\ep^{ijk}$ and $\ep_{ijk0}=\ep_{ijk}$
with $\ep^{123}=\ep_{123}=1$. We also use the covariantly constant
anti-symmetric tensors given by $\h^{ijk}= \ep^{ijk}/\srg$, and
$\h_{ijk}= \ep_{ijk} \srg$. The Ricci tensor is defined by
\bea g^{ij}[\nb_i,~\nb_k]A_j~=~R_k^{~\l} A_{\l}.  
\label{Ricci}
\eea 
For D=3 spaces, the curvature tensor is expressed in terms of the Ricci
tensor by
\bea
R_{ijk\l}~=~\ep_{ijm}\ep_{k\l n}(-\gamma)\left(R^{mn}-\frac{g^{mn}}{2}R\right).
\label{curv}
\eea
Using \bref{Ricci} and imposing the transverse condition,
\bea
\nb^i A_i =0,
\eea
one obtains
\bea
{\t \Delta}^j_{~i} A_j = \nb^{k} \ep_{ki\l} B^{\l}.
\eea
Assuming the presence of the inverse operator, it is solved as
\bea
A_i~=~({\t \Delta}^{-1})_{i}^{~j} \nb^{k} \ep_{kj\l}B^{\l}.
\eea
For a vector $T_i$, we have
\bea
\nb^\l(\t\Delta^{-1})_\l^{~k}T_k&=&(\Delta^{-1})\nb^{~k}T_k,
\nn\\
\h^{ijk}~\nb_j~(\t\Delta^{-1})_k^{~\l}T_\l
&=&(\t\Delta^{-1})^i_{~j}\h^{jk\l}~\nb_k~T_\l,
\label{fm3}
\eea
where the curvature relation \bref{curv} is used to derive the last equation.
The tensor operator $D$ and its inverse $D^{-1}$ defined in the projected
space by 
\bea
D^{ij} &=& \h^{ikj}~\nb_k~=~\nb_k \h^{ikj}
\nn\\
D^{-1}_{i\l}&=&(\t\Delta^{-1})_i^{~k}\nb^j\h_{jk\l}
~=~\h_{ijk}\nb^k(\t\Delta^{-1})^j_{~\l}, 
\eea
which satisfy
\bea
D^{-1}_{im}~D^{mk}~T_k
&=&
[(\t\Delta^{-1})_i^{~k}\nb^\l~-~(\t\Delta^{-1})_i^{~\l}\nb^k~]~\nb_\l~T_k
\nn\\
&=&
[\D_i^{~k}~-~(\t\Delta^{-1})_i^{~\l}\nb_\l~\nb^k]~T_k\equiv~O_i^{~k}(\nb)~T_k,
\eea
and
\bea
\nb^k D_{k\l}T^\l&=&\nb^k 
(\t\Delta^{-1})_k^{~r}\nb^s\ep_{sr\l}~{\srg~}~T^\l
\nn\\
&=&
(\Delta^{-1})\frac{\ep^{sr\l}}{\srg~}(-R^m_{~\l rs})T_m~=~0,
\label{deltrvc}
\eea
where we have used the Bianchi identity in the last line in
\bref{deltrvc}.
The partial integration formula for the inverse Laplacian operator that
the tensor operators read
\bea
\int d^3x\srg~[T^\l~(\t\Delta^{-1})_{\l}^{~k}~S_k]~&=&~
\int d^3x\srg~\left[~\left((\t\Delta^{-1})^k_{~\l}T^\l\right)S_k)\right],
\nn\\
\int d^3 x \srg~~S^k~D^{-1}_{k\l}~T^\l &=& 
\int d^3 x \srg~~T^k~D^{-1}_{k\l}~S^\l,
\nn\\
\int d^3 x \srg~~S_k~D^{k\l}~T_\l &=& 
\int d^3 x \srg~~T_k~D^{k\l}~S_\l.
\label{painteg}
\eea
The second equation is derived, for instance, as
\bea
\int d^3x~\srg~[T^k~D^{-1}_{k\l}~S^\l]
&=&
-~\int d^3x\srg~[\{\nb^j(\t\Delta^{-1})^{k}_{~\l}T^\l\}~
                  \ep_{jkn}\srg~S^n~]
\nn\\
&=&-~\int d^3x[ (\t\Delta^{-1})_m^{~k}({\srg~}{\ep_{k\l r}})~\nb^\l 
T^r]~                 (\srg~S^m)
\nn\\
&=&
+~\int d^3x~\srg~[S^\l~D^{-1}_{\l n}~T^n].
\label{e11}
\eea

\section{Proof of the on-shell relation in curved spaces} 

In this Appendix, we show that the relation 
\bea
\D F_{0i}~=~{\lam}~K_{0i}~+~{\lam} D^{-1}_{ij} 
\left(\frac{\pa_{\mu}{\t {\cal K}}^{\mu j}}{\srg}\right) + \lam 
 O_i^{~k}
\pa_0\left( \frac{D^{-1}_{kj}}{\srg}\right)~\t {\cal K}^{0j}
\label{C1}
\eea
reduces to $\D F_{0i}=\lam K_{0i}$ on the mass shell.  The second term on the
r.h.s. of \bref{C1} is proportional to EOM \bref{eombi}.  One rewrites the third
term by using the projection property 
$\D_i^{~k}=D^{-1}_{im}~D^{mk}+\nb_i(\Delta^{-1})\nb^k$:
\bea
\pa_0\left( \frac{D^{-1}_{ij}}{\srg}\right)~\t {\cal K}^{0j}=
\pa_0\left( \frac{D^{-1}_{ij}}{\srg}\right)~
\left({D^{jk}}\srg~\frac{D^{-1}_{k\l}}{\srg}~+~\nb^j\frac{1}{\Delta}\nb_\l\right){\t {\cal K}^{0\l}},
\label{C2}
\eea
where the last term is proportional the Gauss law.  Define 
\bea
Y_i~\equiv~{D^{-1}_{ij}}\frac{\t {\cal K}^{0j}}{\srg},
\eea
and note that ${\srg}{D^{jk}}$ has no time dependence when it operates
on $Y_k$: 
\bea
{\srg}{D^{jk}}Y_k~=~\ep^{jik}\nb_i Y_k~=~\ep^{jik}\pa_j Y_k.
\eea
Then, it follows that 
\bea
\pa_0 \left(\frac{D^{-1}_{ij}}{\srg}\right){D^{jk}}\srg~Y_k&=&
\pa_0 \left({\D}^{k}_{i}~-~\nb_i\frac{1}{\Delta}\nb^k \right)~Y_{k}
\nn\\
&=&-~\pa_0 \left(\nb_i\frac{1}{\Delta}\nb^k\right)~Y_k 
=~\nb_i\frac{1}{\Delta}~\pa_0 (\nb^k)~Y_k,
\eea
where we have used the transverse condition, $\nb^k Y_k=0$.  The
third term in \bref{C1} becomes then
\bea
O_{i}^{~k}\pa_0\left({D^{-1}_{kj}}\frac{1}{\srg}\right)~\t {\cal K}^{0j}&=&
O_{i}^{~j}\left[-~\nb_j\frac{1}{\Delta}~\pa_t(\nb^k)~\left({D^{-1}_{k\l}}\frac{\t {\cal K}^{0\l}}{\srg}\right)\right]
\nn\\
&~&+~O_{i}^{~j}\left[\pa_0\left(\frac{D^{-1}_{jm}}{\srg}\right)\nb^m\frac{1}{\Delta}(\pa_\l\t
{\cal K}^{0\l})\right].
\eea
Here the first vanishes because of the projection property of
$O_i^{~j}$, and the second term is proportional to the Gauss law. One
then arrives at the on-shell relation, $\D F_{0i}= \lam K_{0i}$.


\end{document}